\documentclass[fleqn,usenatbib]{mnras}
\usepackage{newtxtext,newtxmath}
\usepackage[T1]{fontenc}
\usepackage{float}
\usepackage{stfloats}
\usepackage{graphicx}
\usepackage{subcaption}

\DeclareRobustCommand{\VAN}[3]{#2}
\let\VANthebibliography\thebibliography
\def\thebibliography{\DeclareRobustCommand{\VAN}[3]{##3}\VANthebibliography}
\usepackage{graphicx}
\usepackage{amsmath}

\usepackage[switch]{lineno} 
\usepackage{xcolor}        
\makeatletter
\let\LN@col\@LN@col
\makeatother
\title[PS1-12sk]{The Physical Properties of PS1-12sk and the implications for its Progenitor System}

\author[Kai-Li Mi et al.]{
Kai-Li Mi,$^{1}$
Shan-Qin Wang,$^{1}$ \thanks{E-mail: shanqinwang@gxu.edu.cn}
Wen-Pei Gan,$^{2}$
Qiu-Ping Huang,$^{1}$
Tao Wang$^{3,4}$
and En-Wei Liang$^{1}$
\\
$^{1}$Guangxi Key Laboratory for Relativistic Astrophysics,
School of Physical Science and Technology, Guangxi University, Nanning 530004,
China\\
$^{2}$Nanjing Hopes Technology Co., Ltd. Nanjing, 210000, China\\
$^{3}$Institute for Frontiers in Astronomy and Astrophysics, Beijing Normal University, Beijing, 102206, China\\
$^{4}$School of Physics and Astronomy, Beijing Normal University No.19, Xinjiekouwai St, Haidian District, Beijing, 100875, China
}

\date{Accepted 2026 April 10. Received 2026 April 10; in original form 2026 February 03}
\pubyear{2026}

\begin{document}
\label{firstpage}
\pagerange{\pageref{firstpage}--\pageref{lastpage}}
\maketitle
\begin{abstract}	
PS1-12sk is a type Ibn supernova (SN) found in a host environment
showing no obvious ongoing star formation, which challenges the massive star
explosion scenario. We use the ejecta-circumstellar medium (CSM) interaction (CSI)
and the CSI plus $^{56}$Ni models in the context of double white dwarf (WD)
merger to fit the bolometric light curve (LC) of PS1-12sk,
since the He emission lines at the photospheric
phases indicated the interaction between the SN ejecta and He-rich CSM.
We find that the CSI model failed to explain the LC, but
the CSI plus $^{56}$Ni model can account for the bolometric LC.
The derived masses of the two WDs and $^{56}$Ni are $\sim 0.70 M_\odot$,
$\sim 0.40 M_\odot$, and $\sim 0.09\,M_\odot$, respectively.
The facts that the ejecta mass ($\sim 0.984 M_\odot$) is well
below the Chandrasekhar limit ($\sim 1.4 M_\odot$) and that
the $^{56}$Ni mass is comparable to the $^{56}$Ni yields of the explosions of some sub-Chandrasekhar
explosion models support the scenario that
PS1-12sk might be from a sub-Chandrasekhar explosion induced by the merger of two low-mass WDs.
The derived innermost radius ($\sim 13.81 \times 10^{12}$ cm) and the mass of the CSM ($\sim 0.116 M_\odot$)
disfavor the possibility that the CSM was formed in the merger phase. We suggest that
the flybys before the merger can account for the position and mass of the CSM.
\end{abstract}

\begin{keywords}
circumstellar matter -- supernovae: general -- supernovae: individual: PS1-12sk
\end{keywords}

\section{Introduction}\label{sec:intro}

The interaction between the ejecta of Supernovae (SNe) and the circumstellar medium (CSM)
can enhance the bolometric luminosities \citep{Chevalier1982,Chevalier1994} and produce
narrow and intermediate-width emission lines.
According to the emission line features, the SNe interacting with CSM
can be classified to SNe IIn \citep{Schlegel1990,Filippenko1997},
Ia-CSM \citep{Dilday2012,Silverman2013,Fox2015,Kool2023,Tsalapatas2025},
Ibn \citep{Matheson2001,Foley2007,Pastorello2007,Mattila2008,Pastorello2008a,Pastorello2008b,Pastorello2016,Hosseinzadeh2017},
and Icn \citep{Perley2022,Gal-Yam2022}.

It had been widely believed that the progenitors of type Ia-CSM are
white dwarfs (WDs), while the progenitors of type
Ibn and Icn SNe are massive stars which experience
mass losses via steady radiation-driven stellar winds and/or eruptions
at the final stages of their lives.
\citep{Pastorello2008a,Pastorello2008b,Perley2022,Gal-Yam2022}.

It should be noted that, however, PS1-12sk which is a SN Ibn was an
exception. It was found in a region on the outskirt of an elliptical
galaxy CGCG 208-042 in the galaxy cluster RXC J0844.9+4258 \citep{Sanders2013}.
The redshift ($z$) and the location of PS1-12sk are 0.054 and
$08^{\rm h}\,44^{\rm m}\,54.86^{\rm s}\, +42^\circ\,58'\,16.89''$ (J2000)), respectively.
The analysis \citep{Sanders2013,Hosseinzadeh2019} favored the scenario
that PS1-12sk exploded in an inactive star formation environment.

The host environment of PS1-12sk is reminiscent of
a fraction of calcium-rich SN Ib (e.g.,SN~2005E, \citealt{Perets2010}),
and Ia-CSM SN~2020aeuh (\citealt{Tsalapatas2025}),
which are believed to be resulted from the explosions of WDs in binary
systems.

\cite{Sanders2013} found that the inferred $^{56}$Ni mass exceeds the total ejecta mass
if the observed LC of PS1-12sk were powered by $^{56}$Ni decay and therefore
disfavored the $^{56}$Ni model. Assuming that the bolometric energy of PS1-12sk near its
peak luminosity was provided by the interaction between the SN ejecta and an He-dominated CSM,
\cite{Sanders2013} estimated that the mass of the He-rich CSM and the mass loss rate ${\dot M}$
are $\sim\,0.06\,M_\odot (10,000\,{\rm km\,s}^{-1}/v_s)^2$ and $\sim\,0.01\,M_\odot {\rm yr}^{-1}$
(supposing that the wind velocity $v_w$ is $ 100 \,{\rm km\,s}^{-1}$ and the CSM radius $R$ is $2\times10^{15}$ cm),
respectively. After analyzing different channels of the explosion of PS1-12sk,
\cite{Sanders2013} proposed that the PS1-12sk might be from the merger of two WDs.

However, \cite{Sanders2013} did not perform quantitative modeling for the
LC of PS1-12sk. As pointed by \cite{Sanders2013}, additional theoretical work for
interpreting the power source for PS1-12sk, the pre-explosion
mass-loss properties, as well as the star formation properties of the explosion site is needed.

In this paper, we perform detailed modeling for the LC of PS1-12sk
and discuss its progenitor system.
In Section \ref{sec:LCfits}, we model the bolometric LC of PS1-12sk.
We discuss our results in Section \ref{sec:discussion}, and
draw some conclusions in Section  \ref{sec:conclusion}.
Throughout this paper, the luminosity distance $D_L$ of PS1-12sk
is assumed to be 238 Mpc \citep{Sanders2013}.

\section{Modeling the Bolometric Light Curves of PS1-12sk.}\label{sec:LCfits}

We first apply \texttt{SUPERBOL} \citep{Nicholl2018} to the
UV ({\textit{uvw2, uvm2, uvw1, u, b, and v} of \textit{UVOT} filters),
optical (\textit{U, B, V, g, r, i, z, y}), and NIR (\textit{J, H, K}) photometric data which are taken from \citet{Sanders2013},
to construct the bolometric LC of PS1-12sk.	The synthesized bolometric LC is shown in Figure \ref{fig:inter} and Table \ref{table:inter}. 

\begin{figure}
	\includegraphics[width=\columnwidth]{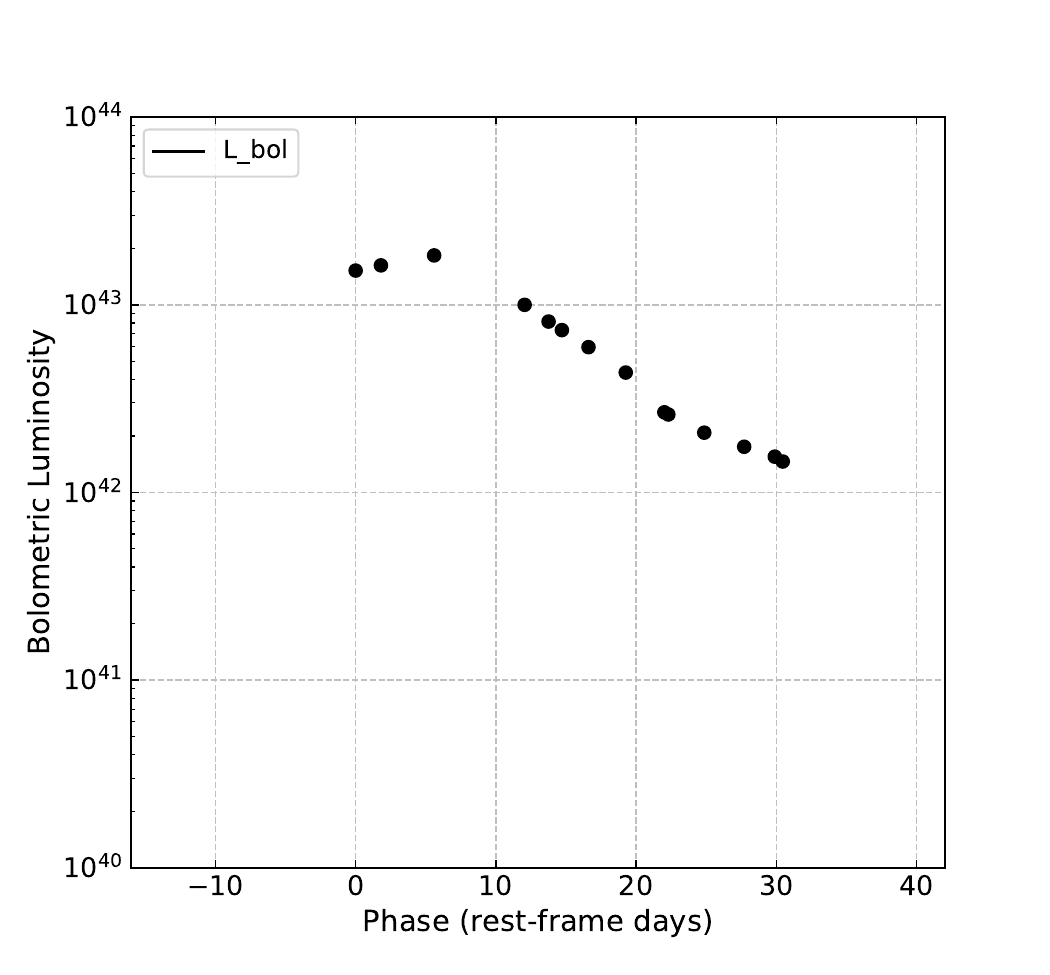}
	\caption{The synthesized bolometric LC of PS1-12sk.}
	\label{fig:inter}
\end{figure}

\begin{table}
	\centering	
	\caption{The derived bolometric luminosity of PS1-12sk at different epochs}
	\begin{tabular}{llllll}
		\hline
		\hline		
		MJD	&Bolometric luminosity  & Error  \\
		& ($10^{42}$ erg s$^{-1}$)  & ($10^{42}$ erg s$^{-1}$)   \\
		\hline
		56000.40                    & 15.2                                               &   0.946             \\
		56002.30                    & 16.2                                               &   0.985             \\
		56006.30                    & 18.3                                               &   1.08              \\
		56013.10                    & 9.98                                               &   0.580             \\
		56014.90                    & 8.13                                               &   0.436             \\
		56015.90                    & 7.32                                               &   0.382             \\
		56017.90                    & 5.94                                               &   0.311             \\
		56020.70                    & 4.35                                               &   0.255             \\
		56023.60                    & 2.67                                               &   0.113             \\
		56023.90                    & 2.60                                               &   0.109             \\
		56026.60                    & 2.08                                               &   0.083             \\
		56029.60                    & 1.75                                               &   0.0722             \\
		56031.90                    & 1.55                                               &   0.0628             \\
		56032.50                    & 1.46                                               &   0.0593             \\
		\hline
	\end{tabular}
	\label{table:inter}
\end{table}
	
There are two possible scenarios which can be expected to produce the explosion of SNe Ibn (including PS1-12sk).
The one assumes that the SN was produced by the merger of the two white dwarfs in a binary system.
Another one supposes that the SN originated from the explosion of a massive star.

Considering the fact that PS1-12sk exploded in a region showing no direct evidence
for star formation \citep{Sanders2013}, we explore the scenario in which a high-mass carbon-oxygen (CO) WD (the primary star)
and a low-mass He WD (the secondary star) merged. The process of the merger can be roughly divided into three steps:
(1) the He WD was disrupted by the tidal force, and a fraction of its mass was stripped in the merger process (e.g., \citealt{Dan2011,Guerrero2004})	and/or expelled by flybys before the merger \citep{Tsalapatas2025},
forming the He-rich CSM;
(2) the rest of the He WD and the CO WD merged and triggered the SN explosion;
(3) the interaction between the SN ejecta and the CSM powered the LC of the SN.
	
We use the CSI and the CSI plus $^{56}$Ni to fit the bolometric LC of PS1-12sk.
The CSI model used here is formulated following \citet{Wang2019}, which is based on \citet{Chevalier1982},
\citet{Chevalier1994}, and \citet{Chatzopoulos2012}, with some revisions. 
The details of the $^{56}$Ni model can be found in \citet{Wang2023}
which is based mainly on \cite{Arnett1982} and \cite{Valenti2008}.
The density profile of the CSM can typically be approximated by a power-law distribution, $\rho_{\rm CSM} = qr^{-s}$,
where $q = \rho_{\rm CSM,in} R_{\rm in}^s$. $R_{\rm in}$ is the innermost radius of the CSM, and $\rho_{\rm CSM,in}$
is the density of the CSM at $R_{\rm in}$. Here $s$ is set to be 2.
{The ejecta is divided into the outer and the inner parts whose density profile 
are $\rho_{\rm ej,outer} \propto r^{-10}$ and $\rho_{\rm ej,inner} \propto r^{-1.1}$.} 
The $\gamma$-ray opacity ($\kappa_\gamma$) is fixed to be 0.027 cm$^2$~g$^{-1}$ (e.g., \citealt{Cappellaro1997}; \citealt{Mazzali2000}; \citealt{Maeda2003}).
The masses of the CO WD and He WD are set to be $M_1$ and $M_2$, respectively.
\footnote{The prior range of the mass of the primary star ($M_1$) is set to be 0$-$1.40 $M_\odot$,
while the prior range of the mass of the donor star ($M_2$) is set be 0.08$-$0.50 $M_\odot$.}
The parameter $a$ is the ratio of the mass becoming the CSM to $M_2$.
Then the ejecta mass ($M_{\rm ej}$) and the CSM mass ($M_{\rm CSM}$) are $M_1 + (1 - a) M_2$ and $a M_2$, respectively.

\begin{figure}
	\centering
	\includegraphics[width=\columnwidth]{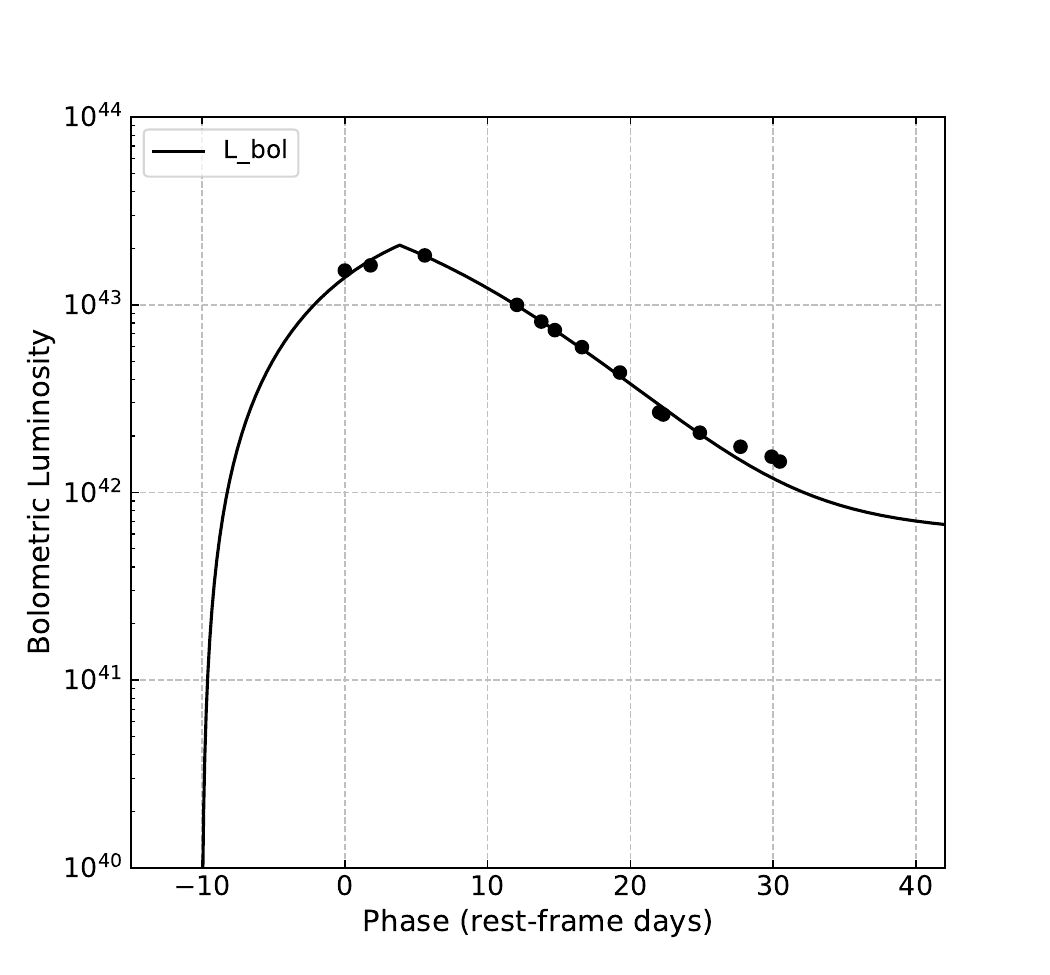}
	\caption{The best fit (solid curves) of the bolometric light curve of PS1-12sk using the CSI model for the white dwarf merger scenario. Shaded regions indicate 1$\sigma$ bounds of the parameters.}
	\label{fig:CSI-WD}
\end{figure}

\begin{figure}
	\centering
	\includegraphics[width=\columnwidth]{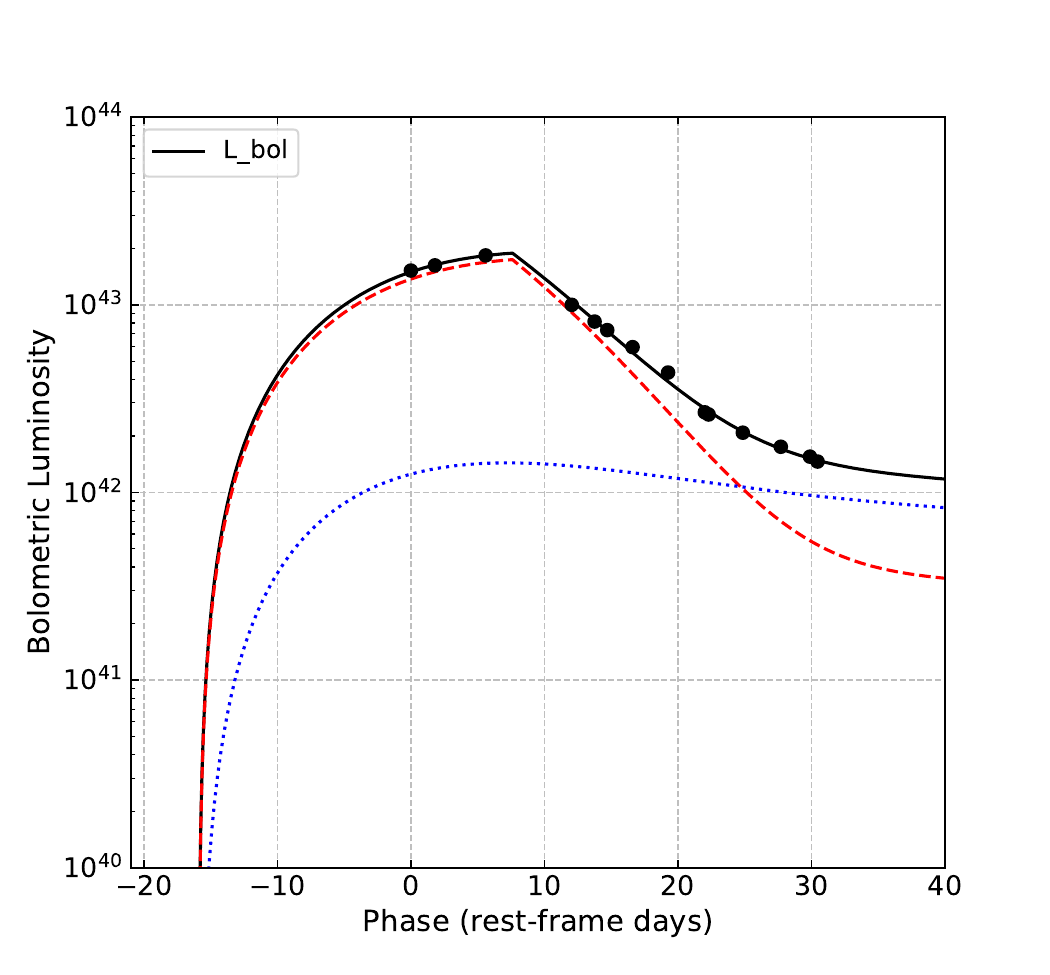}
	\caption{The best fit (the solid curves) of the bolometric LC of PS1-12sk using the CSI plus $^{56}$Ni model (for the WD merger scenario). The dashed and dotted lines represent the contributions from CSI and $^{56}$Ni decay, respectively. The shaded regions indicate 1$\sigma$ bounds of the parameters.}
	\label{fig:Ni-CSI-WD}
\end{figure}
	
We then apply the Markov Chain Monte Carlo (MCMC) method using the Python package \texttt{emcee} \citep{Foreman-Mackey2013}
to derive the best-fitting parameters and their $1\sigma$ uncertainties which are
defined as the 16th and 84th percentiles of the posterior distributions.
We use 20 walkers with 30,000 steps each.
The definitions, the units, and the priors of the parameters of the CSI and the CSI plus $^{56}$Ni
models are listed in Tables \ref{table:CSI-WD} and \ref{table:Ni-CSI-WD}.
\footnote{In our likelihood function, the total variance is modeled as
		\begin{equation}
			\sigma^2 = y_{\rm err}^2 + f^2 \, (\rm{model})^2 .
		\end{equation}
		Here, $f>0$ is a nuisance parameter that quantifies the relative level of systematic uncertainty associated with the model predictions.}
	
The fitting of the LCs of PS1-12sk using the two models are shown in Figures \ref{fig:CSI-WD} and \ref{fig:Ni-CSI-WD}.
The parameters are presented in Table \ref{table:CSI-WD} and Table \ref{table:Ni-CSI-WD}, respectively.
We found that the CSI model cannot fit the late-time bolometric LC. Moreover, the ejecta mass derived by the model is $\sim 1.55M_\odot$,
higher than the Chandrasekhar limit ($\sim 1.4M_\odot$) which would lead to normal SN Ia explosion.

The CSI plus $^{56}$Ni model can fit the whole bolometric LC.
This demonstrated that a moderate of $^{56}$Ni is necessary for powering its
late-time bolometric LC.
The medians, $1\sigma$ bounds, and best-fitting values (which are listed in parentheses) of the parameters are
	$M_{\rm 1}$ = $0.70^{+0.21}_{-0.19}(0.53)$ M$_\odot$,
	$M_{\rm 2}$ = $0.40^{+0.06}_{-0.07}(0.38)$ M$_\odot$,
	$a$ = $0.29^{+0.06}_{-0.06}(0.26)$,
	$M_{\rm Ni}$ = $0.09^{+0.01}_{-0.01}(0.09)$ M$_\odot$,
	$v_{\rm ph}$ = $1.87^{+0.23}_{-0.22}(2.01)$ $\times 10^9$ cm s$^{-1}$,
	$\rho_{\rm CSM,in}$ = $44.50^{+13.02}_{-10.33}(47.62)$ $\times 10^{-12}$ g cm$^{-3}$,
	$R_{\rm in}$ = $13.81^{+2.23}_{-1.73}(12.70)$ $\times 10^{12}$ cm,
	$\epsilon$ = $0.76^{+0.17}_{-0.18}(0.75)$,
	$x_0$ = $0.43^{+0.05}_{-0.06}(0.41)$,
	$ k $ = $0.18^{+0.01}_{-0.02}(0.19)$ cm$^{2}$ g$^{-1}$,
	$t_{\rm shift}$ = $-16.80^{+2.94}_{-3.92}(-15.96)$ days,
	ln $f$ = $-7.02^{+2.33}_{-2.00} (-8.39)$ mag. 
	
We present the corner plot (posteriors) for the fit of the CSI plus $^{56}$Ni model
in Figure \ref{fig:A_Ni_CSI_WD}. It shows that most parameters display unimodal peak distributions
and Gaussian-like profiles. The optical opacity ($\kappa$) favor a value near the upper limit of the prior range,
similar to the cases of SN~2015bn, SN~2010gx, LSQ14mo, LSQ14bdq, PS1-14bj, DES14X3taz, SCP-06F6, and so on
(see Figure 4 and Table 3 of \citealt{Nicholl2017}).
The overall convergence of the MCMC chains indicates that they are well constrained by the data.

\section{Discussion}
\label{sec:discussion}

\subsection{The implications of the derived parameters of PS1-12sk}
\label{subsec:parameter}

According to the values of $M_{\rm 1}$, $M_{\rm 2}$ and $a$ derived by the CSI plus $^{56}$Ni model,
we can infer that $M_{\rm ej}$ of PS1-12sk is $\sim 0.984 M_\odot$, lower
than the Chandrasekhar limit. This indicates that PS1-12sk might be from
a sub-Chandrasekhar explosion of in a WD binary system.
The derived value of $M_{\rm Ni}$ of PS1-12sk is
$\sim 0.09 M_\odot$, which is comparable to the $^{56}$Ni yield of some
sub-Chandrasekhar explosion models \citep{Woosley2011,Taubenberger2017}, further
supporting the sub-Chandrasekhar explosion scenario.

\begin{table*}
	\centering	
	\caption{The Definitions, Units, Prior, Medians, $1\sigma$ Bounds, and best-fitting values of the Parameters of the CSI model.}
	\begin{tabular}{llllll}
		\hline
		\hline		
		Parameters	&Definition  & Unit  & Prior   	& Medians & Best-fitting\\
		&   &    &    	&  &  values\\
		\hline
		$M_{\rm 1}$                    &  The primary star mass                                        &   $M_\odot$             &    $[0.5, 1.4]$    &$1.22^{+0.11}_{-0.13} $& 1.22 \\
		$v_{\rm ph}$                    & {The} early-time photospheric velocity               		   &   {$10^4$ km s$^{-1}$}  &    $[0.1, 4]$    &$3.59^{+0.19}_{-0.27} $& 3.54\\
		$M_{\rm 2}$                    & The companion star mass                                		&   $M_\odot$          	  &	   $[0.3, 0.5]$  	  &$0.35^{+0.05}_{-0.03} $& 0.33\\
		$ a $                          & The CSM mass fraction from the companion  				    	&    -                   &    $[0.03, 0.12]$    &$0.04^{+0.01}_{-0.00} $& 0.03\\
		$\rho_{\rm CSM,in}$            & The innermost CSM density 					 &   $10^{-12}$ g cm$^{-3}$       & $[0.001, 100]$  &$58.89^{+10.59}_{-9.69} $& 52.95\\
		$R_{\rm in}$                   & The innermost radius of the CSM								&   $10^{12}$ cm         &    $[0.01, 2000]$  &$28.47^{+2.86}_{-3.16} $& 29.87\\
		$\epsilon$                      & The conversion efficiency from the kinetic energy to radiation&   -                      &    $[0.01, 0.9]$   &$0.74^{+0.08}_{-0.10} $& 0.75\\
		$x_{\rm 0}$                     & Dimensionless radius                                          &   -                   &    $[0.01, 0.5]$  &$0.27^{+0.03}_{-0.02} $& 0.27\\
		$ k $		                    &  The optical opacity                                         &  cm$^{2}$ g$^{-1}$       &     $[0.05, 0.2]$     &$0.14^{+0.03}_{-0.04} $& 0.15\\	
		$t_{\rm shift}$                 & The explosion time relative to the first data                 &   days                  &    $[-20, 0]$     &$-1.35^{+0.14}_{-0.25} $& $-1.30$\\
		$\ln f$                        & The systematic errors                                                             &  mag           &           $[-10, -3.5]$     &$-3.51^{+0.01}_{-0.01} $& $-3.50$\\
		\hline
	\end{tabular}
	\label{table:CSI-WD}
\end{table*}

\begin{table*}
	\centering	
	\caption{The Definitions, Units, Prior, Medians, $1\sigma$ Bounds, and best-fitting values of the Parameters of the CSI plus $^{56}$Ni model.}
	\begin{tabular}{llllll}
		\hline
		\hline		
		Parameters	&Definition  & Unit  & Prior   	& Medians & Best-fitting\\
		&   &    &    	&  &  values\\
		\hline
		$M_{\rm 1}$                    &  The primary star mass                                        &   $M_\odot$             &    $[0, 1.4]$      &$0.70^{+0.21}_{-0.19}$ & 0.53 \\
		$M_{\rm 2}$                    & The companion star mass                                		&   $M_\odot$          	  &	   $[0.08, 0.5]$  	  &$0.40^{+0.06}_{-0.07}$ & 0.38 \\
		$ a $                          & The CSM mass fraction from the companion  				    	&    -                   &    $[0, 0.5]$    &$0.29^{+0.06}_{-0.06}$ & 0.26 \\
		$M_{\rm Ni}$                    & The Ni mass                                 					&   $M_\odot$          	  &	   $[0.009, 0.2]$   &$0.09^{+0.01}_{-0.01}$ & 0.09 \\
		$v_{\rm ph}$                    & {The} early-time photospheric velocity               		   &   {$10^4$ km s$^{-1}$}  &    $[0.1, 2.2]$    &$1.87^{+0.23}_{-0.22}$ & 2.01 \\
		{$\rho_{\rm CSM,in}$}            & The innermost CSM density 										 &   $10^{-12}$ g cm$^{-3}$      & $[10, 100]$    &$44.50^{+13.02}_{-10.33}$ & 47.62 \\
		$R_{\rm in}$                   & The innermost radius of the CSM								&   $10^{12}$ cm         &    $[0.01, 50]$  &$13.81^{+2.23}_{-1.73}$ & 12.70 \\
		$\epsilon$                      & The conversion efficiency from the kinetic energy to radiation&   -                      &    $[0.01, 1]$   &$0.76^{+0.17}_{-0.18}$ & 0.75 \\
		$x_{\rm 0}$                     & Dimensionless radius                                          &   -                   &    $[0.01, 0.5]$     &$0.43^{+0.05}_{-0.06}$ & 0.41 \\
		$ k $		                    &  The optical opacity                                          &  cm$^{2}$ g$^{-1}$       &     $[0.05, 0.2]$     &$0.18^{+0.01}_{-0.02}$ & 0.19 \\
		$t_{\rm shift}$                 & The explosion time relative to the first data                 &   days                &    $[-25, -10]$     &$-16.80^{+2.94}_{-3.92}$ & $-15.96$ \\
		$\ln f$                        & The systematic errors                                           &  mag                   & $[-10, -3.5]$     &$-7.02^{+2.33}_{-2.00}$ & $-8.39$ \\
		\hline
	\end{tabular}
	\label{table:Ni-CSI-WD}
\end{table*}

It should be noted that some Ca-rich Ib SNe exploding in old environment
have low $^{56}$Ni yield which is $\lesssim 0.1 M_\odot$
(e.g., $M_{\rm Ni}$ of PTF~10iuv is $\sim 0.016 M_\odot$ \citep{Kasliwal2012},
$M_{\rm Ni}$ of SN~2022oqm is $\sim 0.1 M_\odot$ \citep{Irani2024}.)
This feature resembles that of PS1-12sk, suggesting that PS1-12sk
and some Ca-rich SNe might exploded under the same explosion mechanism,
e.g., the detonations or deflagrations associated with low-mass
(i.e., sub-Chandrasekhar) explosion of WDs \citep{Perets2010,De2020,Zenati2023}. 
\footnote{Different values of $s$ would result in different values of $^{56}$Ni mass and all other parameters;  
considering that the sub-Chandrasekhar explosions produce low $^{56}$Ni masses,  
we suggest that our derived $^{56}$Ni mass might be an upper limit.}

The value of $v_{\rm ph}$ was not determined by \cite{Sanders2013}. 
\cite{Sanders2013} inferred that the velocity of the CSM is $\sim 3000 $ km s$^{-1}$ and 
suggest that this value can be set to be the lower limit of $v_{\rm ph}$. 
Our derived $v_{\rm ph}$ ($1.87^{+0.23}_{-0.22}$ $\times 10^9$ cm s$^{-1}$) 
is consistent with this constraint.

\subsection{The origin of the CSM of PS1-12sk}

The derived CSM mass of the CSI plus $^{56}$Ni model is $\sim 0.116 M_\odot$.
Assuming that the CSM was from the tidal tail in
the merger phase, the derived $R_{\rm in} (\sim 13.81 \times 10^{12}$ cm) needs $\gtrsim 10^{4}-10^{5}$ s of
time delay (assuming that the CSM velocity is $\sim 10^{8}-10^{9}$ cm s$^{-1}$).
The large value is significantly longer than the dynamical timescale before detonation
($\sim 10-100$ s \citep{Inoue2025} and references therein).
Besides, the derived value of $a$ ($\sim$ 0.29) is significantly larger
than $\sim$ 0.03 which was derived by \cite{Dan2011} and
$\sim$ 0.10 which was derived by \cite{Guerrero2004}.
These indicate that the CSM cannot be produced in the merger phase.

Alternative mechanisms producing the CSM include super-Eddington winds \citep{Inoue2025}
and flybys of the double WDs before the mergers.
The two scenario has enough time for the material to expand to the distances $\gtrsim 10^{13}$ cm.
For the super-Eddington wind scenario, however, the CSM expelled is
a few $0.01 M_\odot$ with upper limit
$\sim 0.03 M_\odot$ \citep{Inoue2025} which is also lower than the
PS1-12sk's CSM mass derived here. For the flyby scenario,
the CSM can be a few $0.1 M_\odot$ for the CO WD merger \citep{Tsalapatas2025}.
It is reasonable to expect that the stripped mass
can also be $\gtrsim 0.1 M_\odot$ at the flyby phase of CO-He WD system.

Therefore, we suggest that the CSM of PS1-12sk
might be from the flybys before the merger of the
two WDs.

\section{Conclusions}
\label{sec:conclusion}

In this paper, we constrain the physical properties of SN Ibn PS1-12sk
and explore its progenitor system.
We use the CSI model and the CSI plus $^{56}$Ni model in the
context of WD-WD merger to fit its bolometric LC synthesized from
the observed multi-band photometry, since the
host environment of explosion site of PS1-12sk do not show evident
ongoing star formation.

We find that the CSI model cannot fit the late-time LC of PS1-12sk and
the CSI plus $^{56}$Ni model can account for the LC and the parameters
are reasonable. This indicates that the contribution of $^{56}$Ni to
the LC cannot be neglected.
The derived $^{56}$Ni mass is $\sim 0.09\,M_\odot$,
which is comparable to the $^{56}$Ni yields of the explosions of some sub-Chandrasekhar
explosion models, sub-luminous Ia SNe, and some Ca-rich SNe exploded in the host
environment similar to that of PS1-12sk.
The derived masses of the two WDs are $\sim 0.70 M_\odot$ and
$\sim 0.40 M_\odot$, respectively. The ejecta mass is $\sim 0.984 M_\odot$, well
below the Chandrasekhar limit ($\sim 1.4 M_\odot$).
These two features support the scenario in which the mergers of double WD systems trigger
sub-Chandrasekhar explosions.

Our modeling shows that $\sim 0.29$ of the mass of the He WD ($\sim 0.40M_\odot$)
became $\sim 0.116 M_\odot$ of CSM and the innermost radius of the CSM ($R_{\rm in}$)
is $\sim 13.81 \times 10^{12}$ cm).
The value of $a$ ($\sim$ 0.29) is significantly larger than those
obtained by numerical simulations ($\lesssim$0.03 to
$\lesssim$ 0.10) in the context of tidal tail, while the
CSM mass is about one magnitude higher than that
inferred by the super-Eddington wind scenario.
Nevertheless, the CSM mass can be accounted for the
flyby scenario which can shed a few tenths $M_\odot$
of CSM in double CO WD system.

We caution that, while our modeling can reproduce the LC of PS1-12sk
and the derived parameters are reasonable,
it is based on the semi-analytic models 
and we only deal with the case of $s$=2.     
Another caveat of the study is that the X-ray and radio upper limits have not been used to pose more stringent constraints on the  
properties of the CSM.
We suggest that more realistic numerical simulations for the multi-band LCs (specifically of X-ray and radio data) of PS1-12sk and similar events are needed to break possible parameter degeneracy of the models
and better constrain the physical properties and the progenitor systems.

\section*{Acknowledgements}

We thank the anonymous referee for helpful comments and 
suggestions that have allowed us to improve this manuscript. 
We thank Shuo-Jin You and Yun-Feng Liang for helpful discussion. 
This work is supported by the Guangxi Science Foundation 
(grant No. 2025GXNSFDA02850010), the Guangxi Science and Technology Innovation Platform Program (Leitai Action Plan, Grant No. Guike LT2600640026), Guangxi Key R\&D Program (Guangxi Funeng Action Plan, Grant No. Guike FN2504240040), and the "Guangxi Highland of Innovation Talents'' Program. 
This work is also supported by the National Natural Science Foundation of China 
(grant No. 12494571, 12133003, and 11963001) and Program of Bagui Scholars (LHJ).

\section*{Data Availability}

The photometric data (UV, optical, and NIR) of PS1-12sk used in this work are taken from Sanders et al. (2013). The bolometric light curve constructed is listed in Table \ref{table:inter} of this paper.

\bibliographystyle{mnras}
\bibliography{references}

\appendix

\section{Some extra material}

Figures \ref{fig:A_Ni_CSI_WD} presents the corner plot of the CSI plus $^{56}$Ni model.

\begin{figure*}
	\centering
	\includegraphics[width=\textwidth]{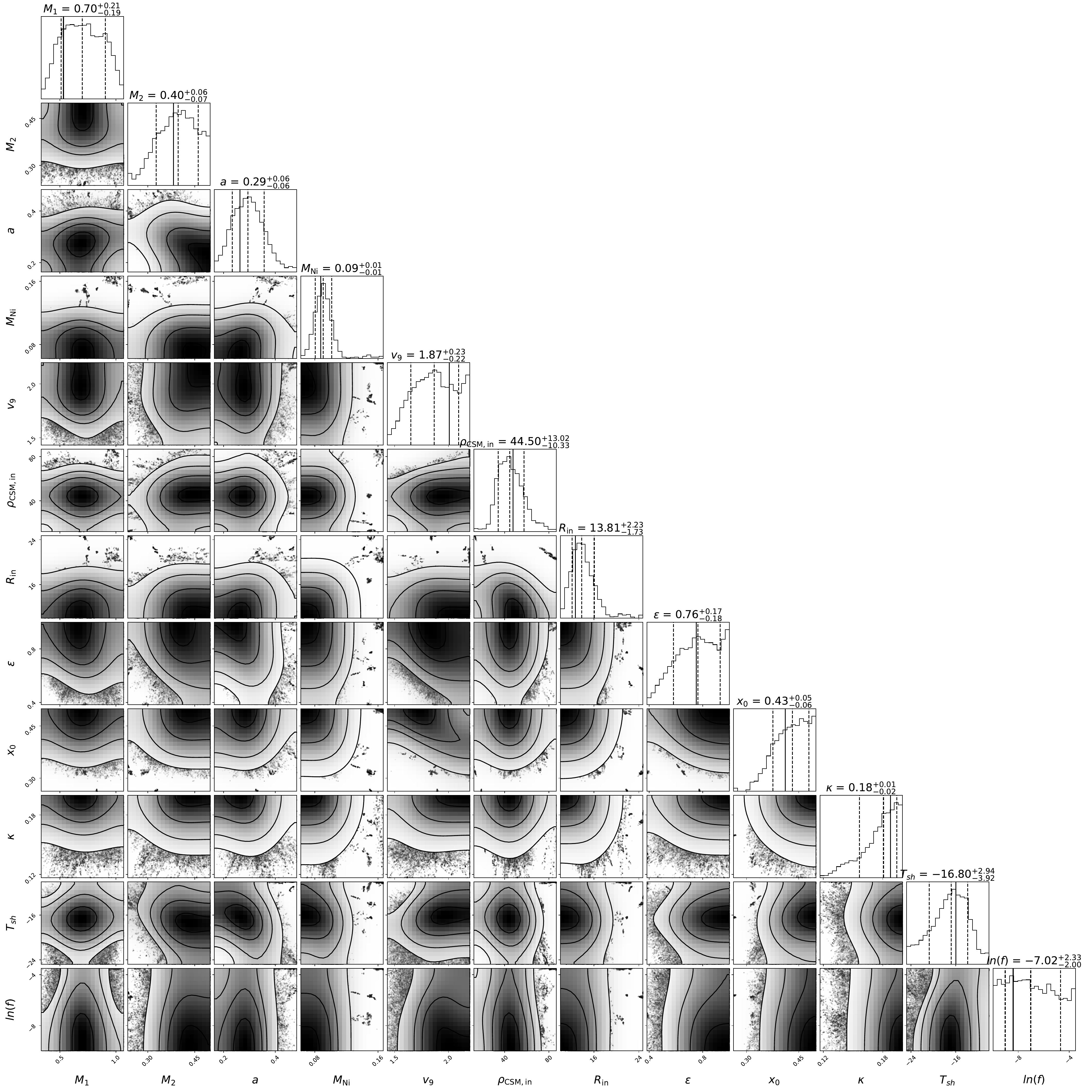}
	\caption{The corner plot of the CSI plus $^{56}$Ni  model. The solid vertical lines represent the best-fitting parameters, while
		the dashed vertical lines represent the medians and the $1\sigma$ bounds of the parameters.
		\label{fig:A_Ni_CSI_WD}}
\end{figure*}

\bsp	
\label{lastpage}
\end{document}